\documentclass[12pt]{article}

\usepackage[dvips]{graphicx}
\usepackage{amsfonts}
\usepackage{amsmath}
\usepackage{amssymb}

\numberwithin{equation}{section}

\newcommand{\bR}{{\mathbb R}}
\newcommand{\bT}{{\mathbb T}}
\newcommand{\bD}{{\mathbb D}}
\newcommand{\bC}{{\mathbb C}}

\newcommand{\bN}{{\mathbb N}}

\newcommand{\kF}{{\mathcal F}}

\newcommand{\kO}{{\mathcal O}}

\newcommand{\gotB}{{\mathfrak B}}

\newcommand{\gotH}{{\mathfrak H}}
\newcommand{\goth}{{\mathfrak h}}

\newcommand{\gotk}{{\mathfrak k}}

\newcommand{\gotX}{{\mathfrak X}}

\newcommand{\ga}{{\alpha}}
\newcommand{\gb}{{\beta}}
\newcommand{\gd}{{\delta}}
\newcommand{\gD}{{\Delta}}
\newcommand{\gga}{{\gamma}}

\newcommand{\gl}{{\lambda}}

\newcommand{\gt}{{\tau}}
\newcommand{\gth}{{\theta}}

\newcommand{\slim}{\,\mbox{\rm s-}\hspace{-2pt} \lim}

\newcommand{\real}{{\Re{\mathrm e\,}}}
\newcommand{\imag}{{\Im{\mathrm m\,}}}
\newcommand{\dom}{{\mathrm{dom}}}

\newtheorem{theo}{Theorem}[section]
\newtheorem{prop}[theo]{Proposition}
\newtheorem{lem}[theo]{Lemma}
\newtheorem{cor}[theo]{Corollary}

\newtheorem{defi}[theo]{Definition}
\newtheorem{rem}[theo]{Remark}

\newcommand{\ba}{\begin{array}}
\newcommand{\ea}{\end{array}}
\newcommand{\bea}{\begin{eqnarray}}
\newcommand{\eea}{\end{eqnarray}}
\newcommand{\bead}{\begin{eqnarray*}}
\newcommand{\eead}{\end{eqnarray*}}
\newcommand{\be}{\begin{equation}}
\newcommand{\ee}{\end{equation}}
\newcommand{\bed}{\begin{displaymath}}
\newcommand{\eed}{\end{displaymath}}
\newcommand{\bl}{\begin{lem}}
\newcommand{\el}{\end{lem}}
\newcommand{\bp}{\begin{prop}}
\newcommand{\ep}{\end{prop}}
\newcommand{\bt}{\begin{theo}}
\newcommand{\et}{\end{theo}}
\newcommand{\Label}{\label}
\newcommand{\bc}{\begin{cor}}
\newcommand{\ec}{\end{cor}}
\newcommand{\la}{\Label}

\newcommand{\br}{\begin{rem}}
\newcommand{\er}{\end{rem}}
\newcommand{\bd}{\begin{defi}}
\newcommand{\ed}{\end{defi}}

\def\wt#1{{{\widetilde #1} }}
\def\wh#1{{{\,\widehat #1\,} }}

\newenvironment{proof}%
{\begin{sloppypar}\noindent{\bf Proof.}}%
{\hspace*{\fill}$\square$\end{sloppypar}\bigskip}

\begin{document}

\title{\vspace{-3.0cm}
Trotter-Kato product formula for unitary groups}

\author{Pavel Exner$^{1,2}$ and Hagen Neidhardt$^3$}

\date{\large 1) Department of Theoretical Physics, NPI\\
Academy of Sciences, \mbox{CZ-25068} \v Re\v z \\
E-mail: exner@ujf.cas.cz\\[1mm]
2) Doppler Institute, Czech Technical University\\\
B\v{r}ehov{\'a}~7, CZ-11519 Prague, Czech Republic \\[1mm]
3) Weierstrass Institute for \\
Applied Analysis and Applications\\
Mohrenstrasse 39, D-10117 Berlin, Germany \\
E-mail: neidhard@wias-berlin.de
}

\maketitle

{\bf Keywords:} Trotter product formula, Trotter-Kato product formula,
unitary groups, Feynman path integrals, holomorphic Kato functions\\

{\bf Subject classification:} Primary 47A55, 47D03, 81Q30, 32A40
\newline Secondary 47B25\\

{\bf Abstract:} Let $A$ and $B$ be non-negative self-adjoint operators in a separable
Hilbert space such that its form sum $C$ is densely defined. It is shown
that the Trotter product formula holds for imaginary times in the
$L^2$-norm, that is, one has
\begin{displaymath}
 \lim_{n\to+\infty}\int^T_0 \left\|\left(e^{-itA/n}e^{-itB/n}\right)^nh -
   e^{-itC}h\right\|^2dt = 0
\end{displaymath}
for any element $h$ of the Hilbert space and any $T > 0$. The result
remains true for the Trotter-Kato product formula
\begin{displaymath}
 \lim_{n\to+\infty}\int^T_0 \left\|\left(f(itA/n)g(itB/n)\right)^nh -
   e^{-itC}h\right\|^2dt = 0
\end{displaymath}
where $f(\cdot)$ and $g(\cdot)$ are so-called holomorphic
Kato functions; we also derive a canonical representation for any function of this class.

\section{Introduction}

The aim of this paper is to prove a Trotter-Kato-type
formula for unitary groups. Apart of a pure mathematical interest
such a product formula can be related to physical problems. In
particular, Trotter formula provides us with a way to define
Feynman path integrals \cite{Ex85,JohLap00} and extending it beyond the
essentially self-adjoint case would allow us to treat in this way
Schr\"odinger operators with a much wider class of potentials.

In order to put our investigation into a proper context let
us describe first the existing related results. Let $-A$ and $-B$
be two generators of contraction semigroups in the Banach space
$\gotX$. In the seminal paper \cite{Tro59} Trotter
proved that if the operator $-C$,
\bed
C := \overline{A + B},
\eed
is the generator of a contraction semigroup in $\gotX$, then the formula
\be\la{1.2} 
e^{-tC} =
\textrm{s\,-\!\!}\lim_{n\to\infty}\left(e^{-tA/n}e^{-tB/n}\right)^n
\ee
holds in $t \in [0,T]$ for any $T > 0$. Formula \eqref{1.2} is
usually called the Trotter or Lie-Trotter product formula.
The result was generalized by Chernoff in \cite{Cher68} as
follows: Let $F(\cdot): \bR_+ \longrightarrow \gotB(\gotX)$ be a
strongly continuous contraction valued function such that $F(0) =
I$ and the strong derivative $F'(0)$ exists and is densely
defined. If $-C$, $C := \overline{F'(0)}$, is the generator of a
$C_0$-contraction semigroup, then the generalized
Lie-Trotter product formula
\be\la{1.2a} 
e^{-tC} =
\textrm{s\,-\!\!}\lim_{n\to\infty}F(t/n)^n 
\ee
holds for $t \ge 0$. In \cite[Theorem 3.1]{Cher74} it is shown
that in fact the convergence in the last formula is
uniform in $t \in [0,T]$ for any $T > 0$. Furthermore, in
\cite[Theorem 1.1]{Cher74} this result was generalized as follows:
Let $F(\cdot): \bR_+ \longrightarrow \gotB(\gotX)$ a family of
linear contractions on a Banach space $\gotX$. Then the
generalized Lie-Trotter product formula \eqref{1.2a} holds
uniformly in $t \in [0,T]$ for any $T > 0$ if and only if there is
a $\gl > 0$ such that
\bed 
(\gl + C)^{-1} = \textrm{s\,-\!\!}\lim_{\gt\to+0}(\gl +
S_\gt)^{-1} 
\eed
where
\bed
S_\gt := \frac{I - F(\gt)}{\gt}, \quad \gt > 0.
\eed
Using the results of Chernoff, Kato was able to prove in
\cite{Ka78} the following theorem: Let $A$ and $B$ be two
non-negative self-adjoint operators in a separable Hilbert space
$\gotH$. Let us assume that the intersection $\dom(A^{1/2}) \cap
\dom(B^{1/2})$ is dense in $\gotH$. If $C := A \stackrel{.}{+} B$
is the form sum of the operators $A$ and $B$, then
Lie-Trotter product formula
\be\la{1.6} 
e^{-tC} = \textrm{s\,-\!\!}\lim_{n\to\infty}
\left(e^{-tA/n}e^{-tB/n}\right)^n 
\ee
holds true uniformly in $t \in [0,T]$ for any $T > 0$. In
addition, it was proven that a symmetrized Lie-Trotter
product formula,
\be\la{1.7} 
e^{-tC} = \textrm{s\,-\!\!}\lim_{n\to\infty}
\left(e^{-tA/2n}e^{-tB/n}e^{-tA/2n}\right)^n, 
\ee
is valid. In fact, the Lie-Trotter formula was extended to
more general products of the form $\left(f(tA/n)g(tB/n)\right)^n$
or $\left(f(tA/n)^{1/2}g(tB/n)f(tA/n)^{1/2}\right)^n$ where $f$
(and similarly $g$) is a real valued function $f(\cdot): \bR_+
\longrightarrow \bR_+$ obeying $0 \le f(t) \le 1$, $f(0) = 1$ and
$f'(0) = -1$ which are called \emph{Kato functions} in the
following. Usually product formul{\ae} of that type are
labeled as Lie-Trotter-Kato.

It is a longstanding open question in linear operator
theory to indicate assumptions under which the
Lie-Trotter product formul{\ae} \eqref{1.6} and \eqref{1.7}
remain to hold for imaginary times, that is, under which
assumptions the formul{\ae}
\be\la{1.3} 
e^{-itC} = \textrm{s\,-\!\!}\lim_{n\to\infty}
\left(e^{-itA/n}e^{-itB/n}\right)^n, \qquad C = A \stackrel{\cdot}{+} B, 
\ee
or
\be\la{1.4} 
e^{-itC} = \textrm{s\,-\!\!}\lim_{n\to\infty}
\left(e^{-itA/2n}e^{-itB/n}e^{-itA/2n}\right)^n, \qquad C = A
\stackrel{.}{+} B, 
\ee
are valid, see \cite[Remark p. 91]{Cher74}, \cite{Far67}, \cite{Ichi80} and \cite{Nel64}. 
We note that if $A$ and $B$ be non-negative selfadjoint 
operators in $\gotH$ and the limit
\bed
U(t) := \slim_{n\to\infty}\left(e^{-itA/n}e^{-itB/n}\right)^n
\eed
exists for all $t \in \bR$, then $\dom(A^{1/2}) \cap \dom(B^{1/2})$
is dense  in $\gotH$ and it holds $U(t) = e^{-itC}$, $t \in \bR$, where $C := A
\stackrel {\cdot}{+} B$, see \cite[Proposition 11.7.3]{JohLap00}. 
Hence it makes sense to assume that $\dom(A^{1/2}) \cap \dom(B^{1/2})$
is dense in $\gotH$. Furthermore, applying Trotter's
result \cite{Tro59} one immediately gets that formul{\ae}
\eqref{1.3} and \eqref{1.4} are valid if $C:= \overline{A+B}$ is
self-adjoint. Modifying Lie-Trotter product formula to a
kind of Lie-Trotter-Kato product formula Lapidus was able
to show in \cite{Lap80}, see also \cite{Lap85}, that one has
\bed
e^{-itC} = \textrm{s\,-\!\!}\lim_{n\to\infty}
\left((I + itA/n)^{-1}(I + itB/n)^{-1}\right)^n
\eed
uniformly in $t$ on bounded subsets of $\bR$. In
\cite{Cach05} Cachia extended the Lapidus result as
follows. Let $f(\cdot)$ be a Kato function which admits a
holomorphic continuation to the right complex plane
$\bC_\mathrm{right} := \{z \in \bC: \real(z) > 0\}$ such
that $|f(z)| \le 1$, $z \in \bC_\mathrm{right}$. Such functions we
call holomorphic Kato functions in the following.
We note that functions from this class admit limits
$f(it) = \lim_{\epsilon\to+0}f(\epsilon + it)$ for a.e. $t \in
\bR$, see Section 5. In \cite{Cach05} it was in
fact shown that if $f$ and $g$ holomorphic Kato functions,
then

\bed
\lim_{n\to\infty}\int^T_0\left\|\left(\frac{f(2itA/n) +
      g(2itB/n)}{2}\right)^nh - e^{-itC}h\right\|^2 dt = 0.
\eed
for any $h \in \gotH$ and $T > 0$. Since $f(t) = e^{-t}$, $t \in
\bR_+$, belongs to the holomorphic Kato class we find
\bed
\lim_{n\to\infty}\int^T_0\left\|\left(\frac{e^{-2itA/n} +
      e^{-2itB/n}}{2}\right)^nh - e^{-itC}h\right\|^2 dt = 0.
\eed
for any $h \in \gotH$ and $T > 0$.

Before we close this introductory survey, let us mention one
more family of related results. The paper \cite{Cach05} was
inspired by a work of Ichinose and one of us \cite{EI05} devoted
to the so-called Zeno product formula which can be regarded
as a kind of degenerated symmetric Lie-Trotter product
formula. Specifically, in this formula one replaces the
unitary factor $e^{-itA/2}$ by an orthogonal projection onto some
closed subspace $\goth \subseteq \gotH$ and defines the operator
$C$ as the self-adjoint operator which corresponds to the
quadratic form $\gotk(h,k) := \left(\sqrt{B}h,\sqrt{B}k\right)$,
$h,k \in \dom(\gotk) := \dom(\sqrt{B})\cap \goth$ where it is
assumed that $ \dom(\gotk)$ is dense in $\goth$. In the
paper \cite{EI05} it was proved that
\bed
\lim_{n\to\infty}\int^T_0 \left\|\left(Pe^{-itB/n}P\right)^nh -
  e^{-itC}h\right\|dt = 0
\eed
holds for any $h \in \goth$ and $T > 0$ where $P$ is the
orthogonal projection from $\gotH$ onto $\goth$.
Subsequently, an attempt was made in \cite{EINZ07} to
replace the strong $L^2$-topology of \cite{EI05} by the usual
strong topology of $\gotH$. To this end a class of
admissible functions was introduced which consisted of
Borel measurable functions $\phi(\cdot): \bR_+ \longrightarrow
\bC$ obeying $|\phi(x)| \le 1$, $x \in \bR_+$, $\phi(0) =1$ and
$\phi'(0) = -i$. It was shown in \cite{EINZ07} that if $\phi$ is
an admissible function such that $\imag(\phi(x)) \le 0$, $x \in
\bR_+$, then
\bed 
e^{-itC} =
\textrm{s\,-\!\!}\lim_{n\to\infty}\left(P\phi(tB/n)P\right)^n
= e^{-itC} 
\eed
holds uniformly in $t \in [0,T]$ for any $T > 0$. We stress
that the function $\phi(x) = e^{-ix}$, $x \in \bR_+$, is
admissible but does not satisfy the condition $\imag(e^{-ix}) \le
0$ for $x \in \bR_+$, and the question about convergence of
the Zeno product formula in the strong topology of $\gotH$ remains
open.

The paper is organized as follows: In Section 2 we formulate our
  main result and relate it to the Feynman integral. In Section 3 is
  devoted to the proof of the main  result. The main result is
  generalized to Trotter-Kato product formulas for 
  holomorphic Kato function in Section 4. Finally, in
  Section 5 we try to characterize holomorphic Kato functions.

\section{The main result}

With the above preliminaries, we can pass to our main result
which can be stated as follows:
\bt\la{I.1}
Let $A$ and $B$ two non-negative self-adjoint operators on the
Hilbert space $\gotH$. If their form sum $C := A
\stackrel{.}{+} B$ is densely defined, then
\be\la{1.16}
\lim_{n\to\infty}
\int^T_0 \left\|\left(e^{-itA/n}e^{-itB/n}\right)^nh -
  e^{-itC}h\right\|^2\;dt =0
\ee
and
\be\la{1.17}
\lim_{n\to\infty}
\int^T_0 \left\|\left(e^{-itA/2n}e^{-itB/n}e^{-itA/2n}\right)^nh -
  e^{-itC}h\right\|^2\;dt = 0
\ee
holds for any $h \in \gotH$ and $T > 0$.
\et
We note that Theorem \ref{I.1} partially solves \cite[Problem
11.3.9]{JohLap00} by changing slightly the topology. 
\br
{\em
From the viewpoint of physical applications, the formula \eqref{1.16} 
allows us to extend the Trotter-type definition of Feynman 
integrals to Schr\"odinger operators with a wider class of potentials.
Following \cite[Definition 11.2.21]{JohLap00} the Feynman integral 
$\kF^t_{\rm TP}(V)$ associated with the potential $V$ is the strong operator limit
\bed
\kF^t_{\rm TP}(V) := \slim_{n\to\infty}\left(e^{-itH_0/n}e^{-itV/n}\right)^n
\eed
where $H_0 := - \tfrac{1}{2}\gD$ and $-\gD$
is the usually defined Laplacian operator in $L^2(\bR^d)$. From \cite[Corollary 11.2.22]{JohLap00}
one gets that the Feynman integral exists if $V: \bR^d \longrightarrow
\bR$ is Lebesgue measurable and non-negative as well as $V \in
L^2_{\rm loc}(\bR^d)$.

Taking into account Theorem \ref{I.1} it is possible to extend the 
Trotter-type definition of Feynman integrals if one replaces the
$L^2(\bR^d)$-topology by the $L^2([0,T]\times\bR^d)$-topology. Indeed,
let us define the generalized Feynman integral $\kF^t_{\rm gTP}(V)$ by
\bed
\lim_{n\to\infty}\int^T_0 
\left\|\left(e^{-itH/n}e^{-itV/n}\right)^nh - \kF^t_{\rm gTP}(V)h\right\|^2 dt = 0
\eed
for $h \in L^2(\bR^d)$ and $T > 0$. Obviously, the
existence of $\kF^t_{\rm TP}(V)$ yields the existence of $\kF^t_{\rm gTP}(V)$
where the converse is in general not true. By Theorem \ref{I.1} 
one immediately gets that the generalized Feynman integral exists if $V: \bR^d
\longrightarrow \bR$ is Lebesgue measurable and non-negative as well as
$V \in L^1_{\rm loc}(\bR^d)$. This essentially extends the class of admissible potentials. 
The same class of potentials is covered by the so-called modified
Feynman integral $\kF^t_M(V)$ defined  by
\bed
\kF^t_M(V) := \slim_{n\to\infty}\left([I + i(t/n)H_0]^{-1}
[I + i(t/n)V]^{-1}\right)^n, 
\eed
see \cite[Definition 11.4.4]{JohLap00} and \cite[Corollary
11.4.5]{JohLap00}.  However, in this case the exponents are replaced by
resolvents which leads to the loss of the typical structure of
Feynman integrals.
}
\er
\br
{\em
\begin{enumerate}

\item[]

\item[\rm (i)]

Formula \eqref{1.16} holds if and only if convergence
in measure takes place, that is, for any $\eta > 0$, $h \in \gotH$ and $T>0$ one has
\be
\lim_{n\to\infty}\left|\left\{t \in [0,T]: \left\|\left(e^{-itA/n}e^{-itB/n}\right)^{n} h -
  e^{-itC}h\right\| \ge \eta\right\}\right| = 0.
\ee
where $|\cdot|$ denotes the Lebesgue measure.

\item[\rm (ii)]
We note that the relation \eqref{1.6} can be rewritten as follows: for any
$\eta > 0$, $h \in \gotH$ and $T > 0$ one has
\bed
\lim_{n\to\infty}\sup_{t \in [0,T]}\left\|\left(e^{-tA/n}e^{-tB/n}\right)^n h -
  e^{-tC}h\right\| = 0.
\eed
This shows that passing to
imaginary times one effectively switches from a uniform
convergence to a convergence in measure.

\item[\rm (iii)]
Theorem \ref{I.1} immediately implies the existence of a
non-decreasing subsequence $n_k \in \bN$, $k \in \bN$, such that
\bed
\lim_{k\to\infty}\left\|\left(e^{-itA/n_k}e^{-itB/n_k}\right)^{n_k} h -
  e^{-itC}h\right\| =0
\eed
holds for any $h \in \gotH$ and a.e. $t \in [0,T]$.
\end{enumerate}
}
\er

\section{Proof of Theorem \ref{I.1}}

The argument is based on the following lemma.
\bl\la{II.1}
Let $\{S_\gt(\cdot)\}_{\gt > 0}$ be a family of bounded
holomorphic operator-valued functions defined in
$\bC_\mathrm{right}$ such that $\real(S_\gt(z)) \ge 0$ for
$z \in \bC_\mathrm{right}$. Let $R_\gt(z) := (I +
S_\gt(z))^{-1}$, $z \in \bC_\mathrm{right}$. If the limit
\bed
\textrm{s\,-\!\!}\lim_{\gt\to+0}R_\gt(t) 
\eed
exists for all $t > 0$, then the following claims are valid:

\item[\rm (i)] The limit
\bed
R(z) := \textrm{s\,-\!\!}\lim_{\gt\to+0}R_\gt(z) 
\eed
exists everywhere in $\bC_\mathrm{right}$, the
convergence is uniform with respect to $z$ in any compact subset
of $\bC_\mathrm{right}$, and the limit function $R(z)$  is
holomorphic in $\bC_\mathrm{right}$.

\item[\rm (ii)] The limits
\bed 
R_\gt(it) := \textrm{s\,-\!\!}\lim_{\epsilon\to+0}
R_\gt(\epsilon + it) 
\eed
and
\bed 
R(it) := \textrm{s\,-\!\!}\lim_{\epsilon\to+0}
R(\epsilon + it) 
\eed
exist for a.e. $t \in \bR$.

\item[\rm (iii)] If, in addition, there is a non-negative self-adjoint
operator $C$ such that the representation $R(t) = (I + tC)^{-1}$
is valid for $t > 0$, then $R(z) = (I + zC)^{-1}$ for $z \in
\overline{\bC_\mathrm{right}}$ and
\be\la{2.5}
\lim_{\gt\to+0}\int^T_0\left\|R_\gt(it)h - (I + itC)^{-1}h\right\|^2 dt = 0
\ee
holds for any $h \in \gotH$ and $T > 0$.
\el
\begin{proof}
The claims (i) and (ii) are obtained easily; the first one is a
consequence of \cite[Theorem 3.14.1]{HPh57}, the second follows
from \cite[Section 5.2]{SzNF70}. It remains to check the
third claim. To prove $R(z) = (I + zC)^{-1}$ we note that $(I +
tC)^{-1}$, $t > 0$, admits an analytic continuation to
$\bC_\mathrm{right}$ which is equal to $(I + zC)^{-1}$, $z
\in \bC_\mathrm{right}$. Since $R(z)$ is an analytic
function in $\bC_\mathrm{right}$, by (i) one immediately
proves $R(z) = (I + zC)^{-1}$ for $z \in \bC_{\rm right}$. In particular, we
get the representation
\bed
R(it) = (I + itC)^{-1}
\eed
for a.e. $t \in \bR$. Furthermore, by \cite[Lemma 2]{Cach05} one
has
\be\la{2.7}
\lim_{\gt\to+0}\int_\bR \left(R_\gt(it)h,v(t)\right)dt =
\int_\bR \left(R(it)h,v(t)\right)dt
\ee
for any $h \in \gotH$ and $v \in L^1(\bR,\gotH)$. Let $p(\cdot)
\in L^1(\bR)$ be real and non-negative, i.e. $p(t) \ge 0$
a.e. in $\bR$. In particular, if $v(t) := p(t) h$ we find
\bed
\lim_{\gt\to+0}\int_\bR p(t) \left(R_\gt(it)h,h\right)dt =
\int_\bR p(t) \left(R(it)h,h\right)dt
\eed
which yields
\be\la{2.10}
\lim_{\gt\to+0}\int_\bR p(t) \real\left\{\left(R_\gt(it)h,h\right)\right\}dt =
\int_\bR p(t) \real\left\{\left(R(it)h,h\right)\right\}dt.
\ee
Since for each $\gt > 0$ the function $S_\gt(z)$ is bounded in
$\bC_\mathrm{right}$ the limit $S_\gt(it) :=
\textrm{s\,-\!}\lim_{\epsilon\to+0} S_\gt(\epsilon + it)$
exists for a.e. $t \in \bR$, see \cite[Section 5.2]{SzNF70}, and
we have $\real(S_\gt(it)) \ge 0$. Furthermore, from \eqref{2.10} we
get
\bea\la{2.10a}
\lefteqn{
\lim_{\gt\to+0}\int_\bR p(t) \left((I + \real\{S_\gt(it)\})
  R_\gt(it)h,R_\gt(it)h\right)dt }\\
& & =\int_\bR p(t)
\real\left\{\left(R_\gt(it)h,h\right)\right\}dt = \int_\bR p(t)
\left\|R(it)h\right\|^2 dt. 
\nonumber \eea
Obviously, we have
\bead
\lefteqn{
\int_\bR p(t) \left\|R_\gt(it)h - R(it)h\right\|^2dt =
\int_\bR p(t)\left\|R_\gt(it)h\right\|^2dt }\\
& & + \int_\bR p(t) \left\|R(it)h\right\|^2dt - 2
\real\left\{\int_\bR p(t) \left(R_\gt(it)h,R(it)h\right)dt\right\}.
\nonumber 
\eead
If $p(t) \ge 0$ for a.e. $t \in \bR$, then
\bead
\lefteqn{
\int_\bR p(t) \left\|R_\gt(it)h - R(it)h\right\|^2dt }\\
& & \le\int_\bR p(t)\left((I +
  \real\{S_\gt(it)\})R_\gt(it)h,R_\gt(it)h\right)dt
\nonumber\\
& & +\int_\bR p(t) \left\|R(it)h\right\|^2dt - 2
\real\left\{\int_\bR p(t) \left(R_\gt(it)h,R(it)h\right)dt\right\}.
\nonumber 
\eead
Choosing $v(t) = p(t)R(it)h$ we obtain from \eqref{2.7} that
\be\la{2.13}
\lim_{\gt\to+0}\int_\bR p(t)\left(R_\gt(it)h,R(it)h\right)dt =
\int_\bR p(t)\left\|R(it)h\right\|^2dt.
\ee
Taking then into account \eqref{2.10a} and \eqref{2.13} we
find
\bed
\lim_{\gt\to+0}\int_\bR p(t) \left\|R_\gt(it)h - R(it)h\right\|^2dt=
0
\eed
and choosing finally $p(t) := \chi_{[0,T]}(t)$, $T > 0$, we
arrive at the formula \eqref{2.5} for any $h \in \gotH$ and $T >
0$.
\end{proof}
Now we are in position to prove Theorem \ref{I.1}. We set
\bed
F_\gt(z): = e^{-\gt zA/2}e^{-\gt zB}e^{-\gt zA/2}, \quad \gt \ge 0,
\eed
and
\bed
S_\gt(z) := \frac{I - F_\gt(z)}{\gt}, \quad \gt > 0,
\eed
for $z \in \overline{\bC_\mathrm{right}}$. Obviously, the
family $\{S_\gt(\cdot)\}_{\gt > 0}$ consists of bounded
holomorphic operator-valued functions defined in
$\bC_\mathrm{right}$. Since $\|F_\gt(z)\| \le 1$ for $z \in
\bC_\mathrm{right}$ we get that $\real\{S_\gt(z)\} \ge 0$
for $z \in \bC_\mathrm{right}$ and $\gt
> 0$. Using formula (2.2) of \cite{Ka78} we find
\bed 
\textrm{s\,-\!\!}\lim_{\gt\to+0}(I + S_\gt(t))^{-1} =
(I + tC)^{-1} 
\eed
for $t \in \bR$. Obviously, we have
\bed
R_\gt(it) = (I + S_\gt(it))^{-1}
\eed
for a.e $t \in \bR$ where
\bed
S_\gt(it) = \frac{I - e^{-i\gt tA/2}e^{-i\gt tB}e^{-i\gt tA/2}}{\gt}
\eed
for $t \in \bR$ and $\gt > 0$. Applying Lemma \ref{II.1} we obtain
\be\la{3.20}
\lim_{\gt\to+0}\int^T_0 \left\|(I + S_\gt(it))^{-1}h - (I +
  itC)^{-1}h\right\|^2 dt = 0
\ee
for any $h \in \gotH$ and $T > 0$.

Now we pass to $\gotH$-valued functions introducing
$\wh{\gotH} :=  L^2([0,T],\gotH)$. We set
\bed
(\wh{A}f)(t) = tAf(t), \quad f \in \dom(\wh{A}) = \{f \in \wh\gotH:
tAf(t) \in \wh\gotH\}
\eed
and in the same way we define $\wh{B}$ and $\wh{C}$
associated with the operators $B$ and $C$, respectively.
It is obvious that the operators $\wh A$, $\wh B$ and $\wh
C$ are non-negative. Setting
\bed
\wh{F}_\gt := e^{-i\gt \wh A/2}e^{-i\gt \wh B}e^{-i\gt \wh A/2}, \quad \gt
> 0,
\eed
and
\bed
\wh S_\gt := \frac{\wh I - \wh F_\gt}{\gt}, \quad \gt > 0,
\eed
we have
\bed 
(\wh{F}_\gt \wh h)(t) = F_\gt(it)\wh h(t) \quad \mbox{and}
\quad (\wh{S}_\gt \wh h)(t) = \frac{I -F_\gt(it)}{\gt}\wh h(t),
\eed
where $\wh h \in \wh \gotH$. From Lemma \ref{II.1} one immediately
gets that
\bed
\lim_{\gt\to+0}\|(\wh I + \wh S_\gt)^{-1}\wh h  - (\wh I + \wh
C)^{-1}\wh h\|_{\wh \gotH} = 0
\eed
for any $\wh h \in \wh \gotH$. Applying now \cite[Theorem
1.1]{Cher74} we find
\bed
 \textrm{s\,-\!\!}\lim_{n\to\infty}\wh F_{s/n}^n =
e^{-is \wh C} 
\eed
uniformly in $s \in [0,\wh T]$ for any $\wh T > 0$ which yields
\bed
\lim_{n\to\infty}\int^T_0
\left\|\left(e^{-istA/2n}e^{-istB/n}e^{-istA/2n}\right)^n\wh h(t) - e^{-ist C}\wh h(t)\right\|^2 dt
= 0
\eed
for any $\wh h \in \wh \gotH$ and $s \in [0,\wh T]$, $\wh T > 0$.
Setting finally $\wh h(t) = \chi_{[0,T]}(t)h$, $h \in
\gotH$, and $s =1$ we arrive at the symmetrized form
\eqref{1.17} of the product formula. To get the other one,
we take into account the relation
\bed
\left(e^{-istA/2n}e^{-itB/n}e^{-itA/2n}\right)^n =
e^{itA/2n}\left(e^{-itA/n}e^{-itB/n}\right)^n e^{-itA/2n}
\eed
which yields
\bead
\lefteqn{
\left\|\left(e^{-itA/2n}e^{-itB/n}e^{-itA/2n}\right)^n h -
  e^{-it C} h \right\|^2 =}\\
& &
\left\|\left(e^{-itA/n}e^{-itB/n}\right)^n e^{-itA/2n}h - e^{-itA/2n}e^{-it C}h\right\|^2
\nonumber
\eead
and through that the sought formula \eqref{1.16}.

\section{A generalization}
Let $f(\cdot)$ be a holomorphic Kato
function. In general, one cannot expect that  for any non-negative operator $A$ the formula
\bed
\textrm{s\,-\!\!}\lim_{\epsilon\to+0}f((\epsilon + it)A) = f(itA)
\eed
would be valid for all $t \in \bR$. This is due to the fact that the limit
$f(iy)$ does not exist for each $y \in \bR_+$, see Section 5. In order to
avoid difficulties we assume in the following that the limit $f(iy)$ exist for all
$y \in \bR$ and indicate in Section~5 conditions which guarantee this
property.
\bt\la{thx01}
Let $A$ and $B$ two non-negative self-adjoint operators on the Hilbert
space $\gotH$. Assume that $C := A \stackrel{.}{+} B$ is densely defined.
If  $f$ and  $g$ be holomorphic Kato functions such that the limit $f(iy)
  = \lim_{x\to+0}f(x+iy)$ exist for all $y \in \bR$, then
\bed
\lim_{n\to\infty}
\int^T_0 \left\|\left(f(itA/n)g(itB/n)\right)^nh -
  e^{-itC}h\right\|^2\;dt =0
\eed
for any $h \in \gotH$ and $T > 0$.
\et
\begin{proof}
We set
\bed
F_\gt(z) := f(\gt zA)g(\gt zB), \quad z \in \bC_\mathrm{right}, \quad \gt \ge 0,
\eed
and
\bed
S_\gt(z) := \frac{I - F_\gt(z)}{\gt}, \quad z \in \bC_\mathrm{right}, \quad
\gt > 0.
\eed
We note that $\{S_\gt(z)\}_{\gt > 0}$ is a family of bounded
holomorphic operator-valued functions defined in $\bC_\mathrm{right}$ obeying
$\real\{S_\gt(z)\} \ge 0$. We set $R_\gt(z) := (I + S_\gt(z))^{-1}$,
$z \in \bC_\mathrm{right}$, $\gt > 0$. By \cite{Ka78} we know that
\bed
\textrm{s\,-\!\!}\lim_{n\to\infty}\left(f(tA/n)g(tB/n)\right)^n = e^{-tC}
\eed
uniformly in $t \in [0,T]$ for any $T > 0$. Applying Theorem 1.1 of
\cite{Cher74} we find
\bed
\textrm{s\,-\!\!}\lim_{\gt\to+0}R_\gt(t) = (I + tC)^{-1}
\eed
for $t \in \bR_+$. Since $S_\gt(z)$, $z \in \bC_\mathrm{right}$, is a
holomorphic continuation of $S_\gt(t)$, $t \in \bR_+$, one gets that
$R_\gt(z)$, $z \in \bC_\mathrm{right}$, is in turn a holomorphic continuation of $R_\gt(t)$, $t \in \bR_+$. Since
\bed
F_\gt(it) := \textrm{s\,-\!\!}\lim_{\epsilon\to+0}F_\gt(\epsilon + it) = f(i\gt
tA)g(i\gt tB), \quad \gt > 0,
\eed
for $t \in \bR$  we find that
\bed
S_\gt(it) := \textrm{s\,-\!\!}\lim_{\epsilon\to+0}S_\gt(\epsilon + it) = \frac{I- f(i\gt
tA)g(i\gt tB)}{\gt}, \quad \gt > 0,
\eed
holds for $t \in \bR$, which further yields
\bed
R_\gt(it) := \textrm{s\,-\!\!}\lim_{\epsilon\to+0}R_\gt(\epsilon + it) = (I + S_\gt(it))^{-1}, \quad \gt > 0,
\eed
for $t \in \bR$. Applying Lemma \ref{II.1} we prove \eqref{3.20}.
Following now the line of reasoning used after formula \eqref{3.20} we
complete the proof.
\end{proof}
Obviously, the Kato functions $f_k(x) := (1 + x/k)^{-k}$, $x \in
\bR_+$, are holomorphic Kato functions. Indeed, each function $f_k$
admits a holomorphic continuation, $f(z) = (1 + z/k)^{-k}$ on $z \in
\bC_\mathrm{right}$and, moreover, the limit
\bed
f_k(it) := \lim_{\epsilon\to+0}f(\epsilon + it) = (1 + it/k)^{-k}
\eed
exists for any $t \in \bR$. This yields
\bed
\lim_{n\to+\infty}\int^T_0 \left\|\left((I + itA/kn)^{-k}(I +
    itB/kn)^{-k}\right)^nh - e^{-itC}h\right\|dt = 0
\eed
for any $h \in \gotH$ and $T > 0$. 
We note that for the particular case  $k=1$ Lapidus demonstrated in
\cite{Lap80} that
\be\la{4.1} 
\slim_{n\to+\infty}\left((I + itA/n)^{-1}(I + itB/n)^{-1}\right)^n =
e^{-itC}
\ee
holds uniformly in $t \in [0,T]$ for any $T > 0$.
By Theorem \ref{thx01} one gets that formula \eqref{4.1} is valid
in a weaker topology as in \cite{Lap80}. This discrepancy
will be clarified in a forthcoming paper.

\section{Holomorphic Kato functions}

\subsection{Representation}

To make use of the results of the previous section one should know 
properties of holomorphic Kato functions. To this purpose we will try 
in the following to find a canonical representation for this function class.
\bt\la{V.1}
If $f$ is a holomorphic Kato function, then
\item[\;\;\rm{(i)}] there is an at most countable set of complex numbers
$\{\xi_k\}_k$, $\xi_k \in \bC_\mathrm{right}$ with $\imag(\xi_k) \ge 0$ satisfying the condition
\be\la{5.1}
\varkappa := 4\sum_k\frac{\real(\xi_k)}{|\xi_k|^2} \le 1
\ee

\item[\;\;\rm{(ii)}]
there is a Borel measure $\nu$ defined on $\overline{\bR}_+ =
[0,\infty)$ obeying $\nu(\{0\}) = 0$ and
\bed
\int_{\bR_+} \frac{1}{1+t^2}\,d\nu(t) < \infty
\eed
such that the limit $\gb := \lim_{x\to+0}\tfrac{2}{\pi}\int_{\bR_+}
\frac{1}{x^2+t^2}\,d\nu(t)$ exists and satisfies the condition
$\gb \le 1 - \varkappa$;

\item[\;\;\rm{(iii)}] the Kato function $f$ admits the representation
\be\la{5.5}
f(x) = D(x)\exp\left\{-\frac{2x}{\pi}\int_{\bR_+}\frac{1}{x^2+t^2}d\nu(t)\right\}e^{-\ga x},
\quad x \in \bR_+,
\ee
where $\ga := 1 -\varkappa - \gb$ and $D(x)$ is a Blaschke-type product given by
\be\la{5.26}
D(x) :=  \prod_k\frac{x^2 - 2x\real(\xi_k) +|\xi_k|^2}{x^2 + 2x\real(\xi_k) + |\xi_k|^2},
\quad  x \in \bR_+.
\ee
The factor $D(x)$ is absent if the set $\{\xi_k\}_k$ is empty; in that case we set $\varkappa := 0$.

Conversely, if a real function $f$ admits the representation
\eqref{5.5} such that the assumptions {\rm (i)} and {\rm (ii)} are
satisfied as well as  $\ga + \varkappa + \gb =1$ holds, then $f$ is a holomorphic Kato function and its
holomorphic extension to $\bC_\mathrm{right}$ is given by
\bed
f(z) = D(z)\exp\left\{-\frac{2z}{\pi}\int_{\bR_+}\frac{1}{z^2+t^2}d\nu(t)\right\}e^{-\ga z},
\quad z \in \bC_\mathrm{right}.
\eed
\et
\begin{proof}
If $f$ is a holomorphic Kato function, then $G(z) := f(-iz)$, $z \in
\bC_+$, belongs to $H^\infty(\bC_+)$. We have $f(z) = G(iz)$,
$z \in \bC_\mathrm{right}$, and taking into account Section C of
\cite{Koo80} we find that if $G(\cdot) \in H^\infty(\bC_+)$, then there is
a real number $\gga \in [0,2\pi)$, a sequence of complex numbers
$\{z_k\}_k$, $z_k \in \bC_+$, satisfying
\be\la{5.6a}
\sum^n_{k=1}\frac{\imag(z_k)}{|i + z_k|^2} < \infty,
\ee
a Borel measure $\nu$ defined on $\bR$ such that
\bed
\int_\bR \frac{1}{1+t^2}\,d\nu(t) < \infty,
\eed
and a real number $\ga \ge 0$ such that $G(\cdot)$ admits the factorization
\bed
G(z) = e^{i\gga}B(z)\exp\left\{-\frac{i}{\pi}\int_\bR
  \left(\frac{1}{z-t} + \frac{t}{1+t^2}\right)d\nu(t)\right\}e^{iaz}, \quad z \in \bC_+,
\eed
where $B(z)$ is the Blaschke product given by
\bed
B(z) := \prod_k \left(e^{i\ga_k}\frac{z -z_k}{z -
    \overline{z_k}}\right),
\quad z \in \bC_+,
\eed
and $\{\ga_k\}_k$ is a sequence of real numbers $\ga_k \in [0,2\pi)$
determined by the requirement
\bed
e^{i\ga_k}\frac{i - z_k}{i - \overline{z_k}} \ge 0.
\eed
The sequence $\{z_k\}_k$ coincides with the zeros of $G(z)$
counting multiplicities.
The quantities $\gga$, $\{z_k\}_k$, $\nu$, $a$ are uniquely determined by $G(\cdot)$.

Using the relation $f(z) = G(iz)$, $z \in \bC_\mathrm{right}$, one gets 
from here a factorization of the holomorphic Kato function,
\be\la{5.11}
f(z) = e^{i\gga}B(iz)\exp\left\{-\frac{i}{\pi}\int_\bR
  \left(\frac{1}{iz-t} + \frac{t}{1+t^2}\right)d\nu(t)\right\}e^{-\ga z},
\ee
$z \in \bC_\mathrm{right}$. Setting next 
$\xi_k = -iz_k \in \bC_\mathrm{right}$ the condition \eqref{5.6a}
takes the form
\bed
\sum^n_{k=1} \frac{\real(\xi_k)}{|1 + \xi_k|^2} < \infty
\eed
and the Blaschke product can be written as
\be\la{5.13}
D(z) := B(iz) = \prod_k \left(e^{i\ga_k}\frac{z -\xi_k}{z + \overline{\xi_k}}\right),
\quad z \in \bC_\mathrm{right},
\ee
where the sequence of real numbers $\{\ga_k\}_k$ is determined now by
\be\la{5.20}
e^{i\ga_k}\frac{1 - \xi_k}{1 + \overline{\xi_k}} \ge 0.
\ee
The complex numbers $\xi_k$ are the zeros of $f(\cdot)$.

Since the Kato function has to be real on $\bR_+$ we easily find that
the condition $f(z) = \overline{f(\overline{z})}$, $z \in \bC_\mathrm{right}$,
has to be satisfied. Hence $\xi_k$ and $\overline{\xi_k}$ are simultaneously zeros of
$f(z)$ and the Blaschke-type product $D(z)$ always contains the factors
$e^{i\ga_k}\frac{z - \xi_k}{z - \overline{\xi_k}}$ and
$e^{-i\ga_k}\frac{z - \overline{\xi_k}}{z - \xi_k}$ simultaneously. 
This allows us to put $D(z)$ into the form
\be\la{5.21}
D(z) =
\prod_k\frac{z^2 - 2z\real(\xi_k) +|\xi_k|^2}{z^2 + 2z\real(\xi_k) + |\xi_k|^2}
\prod_l\frac{z - \eta_l}{z + \eta_l},
\quad  z \in \bC_\mathrm{right},
\ee
where $\real(\xi_k) > 0$, $\imag(\xi_k) > 0$ for complex conjugated
pairs and $\eta_l > 0$ for the remaining real zeros.
Hence we have $D(z) = \overline{D(\overline{z})}$ for $z \in \bC_{\rm right}$. 
Using this relation we find that
\bed
e^{i\gga-g(z)} = e^{-i\gga -\wt{g}(z)}, \quad z \in \bC_\mathrm{right},
\eed
for $z \in \bC_{\rm right}$ where
\bed
g(z) := \frac{i}{\pi}\int_\bR \frac{1 + izt}{iz-t}\,d\mu(t)
\quad \mbox{and} \quad
\wt{g}(z) := \overline{g(\overline{z})} = \frac{i}{\pi}\int_\bR \frac{1 - izt}{iz+t}\,d\mu(t)
\eed
and $d\mu(t) = (1+t^2)^{-1}d\nu(t)$.
Since $g(1) = \wt{g}(1)$ we find $e^{2i\gga} = 1$ which yields $\gga =
0$ or $\gga = \pi$. In both cases we have 
\bed
e^{-g(z)} = e^{-\wt{g}(z)}, \quad z \in \bC_\mathrm{right}.
\eed
By $g(1) = \wt{g}(1)$ we find that $g(z) = \wt{g}(z)$, $z \in
\bC_\mathrm{right}$. Setting $\wt{\mu}(X) := \mu(-X)$ for any Borel set $X$
of $\bR$ we find
\bed
\int_\bR \frac{1 + izt}{iz-t}\,d\mu(t) =
\int_\bR \frac{1 + izt}{iz-t}\,d\wt{\mu}(t),
\quad z \in \bC_\mathrm{right}.
\eed
Using
\bed
\int_\bR \frac{1 + izt}{iz-t}\,d\mu(t) =
(1-z^2)\int_\bR \frac{1}{iz-t}\,d\mu(t)
- \int_\bR d\mu(t)
\eed
and
\bed
\int_\bR \frac{1 + izt}{iz-t}\,d\wt{\mu}(t) =
(1-z^2)\int_\bR \frac{1}{iz-t}\,d\wt{\mu}(t)
- \int_\bR d\wt{\mu}(t)
\eed
as well as the relation $\int_\bR d\mu(t) = \int_\bR d\wt{\mu}(t)$
we find
\bed
\int_\bR \frac{1}{z-t}\,d\mu(t) =
\int_\bR \frac{1}{z-t}\,d\wt{\mu}(t),
\quad z \in \bC_{\rm right},
\eed
which yields $\mu = \wt{\mu}$. Hence the Borel measure obeys
$\mu(X) = \mu(-X)$ for any Borel set $X \subseteq \bR$ 
and this in turn implies $\nu(X) = \nu(-X)$ for any Borel set. Using this property we get
\bead
\lefteqn{
\int_\bR \left(\frac{1}{iz-t} + \frac{t}{1+t^2}\right)d\nu(t) =
\int_\bR \frac{1 +izt}{iz-t}\,d\mu(t) }\\
& &
=\frac{1}{iz}\mu(\{0\}) + \int_{\bR_+}\left(\frac{1 +izt}{iz-t} +
  \frac{1 -izt}{iz+t}\right)d\mu(t),
\quad z \in \bC_\mathrm{right},
\nonumber
\eead
where $\bR_+ = (0,\infty)$. In this way we find
\bed
\int_\bR \left(\frac{1}{iz-t} + \frac{t}{1+t^2}\right)d\nu(t) =
\frac{1}{iz}\nu(\{0\}) -2iz\int_{\bR_+}\frac{1}{z^2 + t^2}d\nu(t)
\eed
for $z \in \bC_\mathrm{right}$.
Summing up we find that a holomorphic Kato function admits the
representation
\bed
f(x) = e^{i\gga}D(x)\exp\left\{-\frac{1}{\pi x}\nu(\{0\}) -
\frac{2x}{\pi}\int_{\bR_+}\frac{1}{x^2+t^2}\,d\nu(t)\right\}e^{-\ga x},
\eed
$x \in \bR_+$, where $D(z)$ is given by \eqref{5.21}.
Since $f(x) \ge 0$, $x \in \bR_+$, one gets that $\gga = 0$ and $D(x) \ge 0$, $x \in
\bR_+$, which means that the real zeros of $f(z)$ are of even
multiplicity. Consequently, the Blaschke-type product $D(z)$ is of the form
\bed
D(z) =  \prod_k\frac{z^2 - 2z\real(\xi_k) +|\xi_k|^2}{z^2 + 2z\real(\xi_k) + |\xi_k|^2},
\quad  z \in \bC_\mathrm{right}.
\eed
We note that the inequality $0 \le f(x) \le 1$, $x \in \bR_+$, is
valid.

Next we have to satisfy the conditions $f(0) := \lim_{x\to+0}f(x) =1$
and $f'(0) = \lim_{x\to+0}\frac{f(x) -1}{x} = -1$. Firstly we note that
\bed
f(x) \le \exp\left\{-\frac{\nu(\{0\})}{\pi x}\right\}, \quad x \in
\bR_+.
\eed
If $\nu(\{0\}) \not= 0$, then it follows that $f(0)=0$ which
contradicts the assumption $f(0) =1$, hence $\nu(\{0\}) = 0$. Next we set
$D_k(x) := \frac{x^2 - 2x\real(\xi_k) +|\xi_k|^2}{x^2 + 2x\real(\xi_k) + |\xi_k|^2}$,
$x \in \bR_+$. Since $0 \le D_k(x) \le 1$, $x \in \bR_+$, we get
\bead
\lefteqn{\hspace{-3mm}
1 - f(x) \ge 1 - D_1(x) + D_1(1 - D_2(x)) + D_1(x)D_2(x)(1 - D_3(x)) + \cdots} \\
& &
+\prod^n_{k=1}D_k(x)\left(1 - \prod_{k=n+1}D_k(x)
\exp\left\{-\frac{2x}{\pi}\int_{\bR_+}\frac{1}{x^2+t^2}\,d\nu(t)\right\}e^{-\ga x}\right)
\nonumber
\eead
for $x \in \bR_+$ and $n =1,2,\ldots$\;. In this way we find the estimate
\bead
\lefteqn{
1 - f(x) \ge 1 - D_1(x) + D_1(x)(1 - D_2(x)) \;+}\\
& &
D_1(x)D_2(x)(1 - D_3(x)) + \cdots  + \prod^{n-1}_{k=1}D_k(x)(1 - D_n(x))
\nonumber
\eead
for $x \in \bR_+$ and $n = 1,2\ldots$\;. This yields
\bead
\lefteqn{
\frac{1 - f(x)}{x} \ge \frac{1 - D_1(x)}{x} + D_1(x)\frac{1 - D_2(x)}{x} \;+}\\
& &
D_1(x)D_2(x)\frac{1 - D_3(x)}{x} + \cdots  + \prod^{n-1}_{k=1}D_k(x)\frac{1 - D_n(x)}{x}
\nonumber
\eead
for $x \in \bR_+$ and $n = 1,2\ldots$\;, and since $\lim_{x\to+0}D_k(x) =
1$ and
\bed
\lim_{x\to+0}\frac{1 - D_k(x)}{x} = 4\frac{\real(\xi_k)}{|\xi_k|^2}
\eed
for $k =1,2,\ldots$\;, we immediately obtain \eqref{5.1}. In
particular, we infer that the limit $D'(0) := \lim_{x\to+0}\frac{D(x) - 1}{x}
= -\varkappa$ exists. Furthermore, we note that condition \eqref{5.1} implies \eqref{5.13}. Furthermore, we have
\bed
1 - f(x) \ge
1 - \exp\left\{-\frac{2x}{\pi}\int_{\bR_+}\frac{1}{x^2+t^2}\,d\nu(t)\right\},
\quad x\in
\bR_+,
\eed
which yields
\bed
\lim_{x\to+0}\exp\left\{-\frac{2x}{\pi}\int_{\bR_+}\frac{1}{x^2+t^2}\,d\nu(t)\right\}
= 1\,,
\eed
or
\bed
\lim_{x\to+0}\frac{2x}{\pi}\int_{\bR_+}\frac{1}{x^2+t^2}\,d\nu(t) = 0.
\eed
Moreover, we have
\bead
\lefteqn{
\frac{1 - f(x)}{x} }\\
& &
\ge \exp\left\{-\frac{2x}{\pi}\int_{\bR_+}\frac{1}{x^2+t^2}\,d\nu(t)\right\}
\frac{\exp\left\{\frac{2x}{\pi}\int_{\bR_+}\frac{1}{x^2+t^2}\,d\nu(t)\right\}
  -1}{x} 
\nonumber\\
& &
\ge\exp\left\{-\frac{2x}{\pi}\int_{\bR_+}\frac{1}{x^2+t^2}\,d\nu(t)\right\}
\frac{2}{\pi}\int_{\bR_+}\frac{1}{x^2+t^2}\,d\nu(t)
\nonumber
\eead
which yields $1 \ge \lim\sup_{x\to+0}\frac{2}{\pi}\int_{\bR_+}\frac{1}{x^2+t^2}\,d\nu(t)$.
However, the function $p(x) :=
\frac{2}{\pi}\int_{\bR_+}\frac{1}{x^2+t^2}\,d\nu(t)$, $x \in \bR_+$, is decreasing which implies the existence of
$\gb :=
\lim_{x\to+0}\frac{2}{\pi}\int_{\bR_+}\frac{1}{x^2+t^2}\,d\nu(t)$.
Summing up these considerations we have found
\be\la{5.36}
f'(0) = \lim_{x\to+0}\frac{f(x)-1}{x} = -\varkappa -\gb -\ga = -1
\ee
which completes the proof of the necessity of the conditions. The
converse is obvious. 
\end{proof}

\subsection{On the existence of $f(iy)$ everywhere}

Besides the fact that $f(x)$ has to be a holomorphic Kato function one
needs that the limit $f(iy) := \lim_{x\to+0}f(x + iy)$
exist for all $y \in \bR$.  First we note that the limit $f(iy)$
exists for a.e. $y \in \bR$. This is a simple consequence of the fact
that the function $G(z) := f(-iz)$, $z \in \bC_\mathrm{right}$, belongs to $H^\infty(\bC_+)$:
for such functions the limit $G(x) := \lim_{\epsilon\to+0}G(x +
i\epsilon)$ exists for a.e. $x \in \bR$ which immediately yields that
$f(iy)$ exists for a.e. $y \in \bR$. To begin with, let us ask about the existence of the limit $|f|(iy) := \lim_{x\to+0}|f(x+iy)|$. For this
purpose we note that the measure $\nu$ of Theorem \ref{V.1} admits
the unique decomposition $\nu = \nu_s + \nu_{ac}$ where $\nu_s$ is
singular and $\nu_{ac}$ is absolutely continuous, and furthermore, the measure $\nu_{ac}(\cdot)$ can be represented as
\bed
d\nu_{ac}(t) = h(t)dt
\eed
where the function $h(t)$ is non-negative and obeys
\bed
\int_{\bR_+}h(t)\,\frac{dt}{1+t^2} < \infty.
\eed
\bp\la{V.2}
Let $f(\cdot)$ be a holomorphic Kato function and let $\gD$ be an open
interval of $\bR$. The limit
$|f|(iy) = \lim_{x\to+0}|f(x + iy)|$ exists for every $y \in \gD$, is
continuous and different from zero on $\gD$ if and only if the limit
\be\la{5.47}
\lim_{x\to+0}|D(x+ iy)| = 1
\ee
exist for every $y \in \gD$, $\nu_s(\gD) = 0$ and the extended weight function
$\wt{h}(t) := h(|t|)$, $t \in \bR$,  is continuous on $\gD$.

In particular, the limit $|f|(iy)$ exists for
every $y \in \bR$, is continuous and different from zero on $\bR$ if
and only if the limit \eqref{5.47} exists for every $y\in\bR$, $\nu_s
\equiv 0$ and the extended function $\wt{h}(\cdot)$ is continuous on $\bR$.
\ep
\begin{proof}
The measure $\nu$ of Theorem \ref{V.1} is given on $[0,\infty)$. We extend it 
to the real axis $\bR$ setting $\nu(X) := \nu(-X)$ for any Borel set
$X \subseteq (-\infty,0)$. Using $\nu(X) := \nu(-X)$ we obtain from
\eqref{5.11} and \eqref{5.13} the representation
\bed
|f(x + iy)| = |D(x + iy)| \exp\left\{-\frac{1}{\pi}
\int_\bR \frac{x}{x^2 + (y+t)^2}\,d\nu(t)\right\}e^{-ax},
\eed
$z = x + iy \in \bC_\mathrm{right}$, or
\bed
|f(x + iy)| = |D(x + iy)| \exp\left\{-\frac{1}{\pi}
\int_\bR \frac{x}{x^2 + (y-t)^2}\,d\nu(t)\right\}e^{-ax},
\eed
$z = x + iy \in \bC_\mathrm{right}$; in this way we find
\bed
-\log(|f(x+iy)|) = -\log(|D(x+iy)|) + \frac{1}{\pi}\int_\bR \frac{x}{x^2 + (y-t)^2}\,d\nu(t) + \ga x
\eed
for $z = x + iy \in \bC_\mathrm{rigth}$. Since
one has $\lim_{x\to+0}|D(x+iy)| =1$ for a.e. $y \in \bR$  we infer that
\bed
-\lim_{x\to+0}\log(|f(x+iy)|) = \lim_{x\to+0}\frac{1}{\pi}\int_\bR \frac{x}{x^2 + (y-t)^2}\,d\nu(t)
\eed
for a.e. $y \in \bR$. Since
\bed
\lim_{x\to+0}\frac{1}{\pi}\int_\bR \frac{x}{x^2 + (y-t)^2}\,d\nu(t) = \wt{h}(y)
\eed
holds for almost all $y \in \bR$ we obtain
$-\log(|f|(iy)) = \wt{h}(y)$ for a.e. $y \in \bR$. By assumption $|f|(iy)$
is continuous and different from zero on $\gD$. Hence the extended weight 
function $\wt{h}(y)$ can be assumed to be continuous on $\gD$.
However, if $\wt{h}(\cdot)$ is continuous on $\gD$, then one has
\bed
\lim_{x\to+0}\frac{1}{\pi}\int_\bR \frac{x}{x^2 + (y-t)^2}\,\wt{h}(t)\,dt = \wt{h}(y)
\eed
for each $y \in \gD$ which means that
\bed
\lim_{x\to+0}\left\{-\log(|D(x+iy)|) +
\frac{1}{\pi}\int_\bR \frac{x}{x^2 + (y-t)^2}\,d\nu_{s}(t)\right\} = 0
\eed
for each $y \in \gD$. Since $ -\log(|D(x+iy)|) \ge 0$ we find
$\lim_{x\to+0}\log(|D(x+iy)|) = 0$ and
%
\bed
\lim_{x\to+0}\frac{1}{\pi}\int_\bR \frac{x}{x^2 + (y-t)^2}\,d\nu_{s}(t) = 0
\eed
for each $y \in \gD$. Taking into account  \cite{Loo43} one can conclude  that the symmetric derivative $\nu'_s(y)$,
\bed
\nu'_s(y) :=
\lim_{\epsilon}\frac{\nu_s((y-\epsilon,y+\epsilon))}{2\epsilon}
\eed
exists and obeys $\nu'_s(y) = 0$ for every $y \in \gD$. If
$\nu_s(\{y_0\}) > 0$ for $y_0 \in \gD$, then
\bed
0 =
\lim_{\epsilon\to+0}\frac{\nu_s((y_0-\epsilon,y_0+\epsilon))}{2\epsilon}
\ge \lim_{\epsilon\to+0}\frac{\nu_s(\{y_0\})}{2\epsilon}
\eed
which yields $\nu_s(\{y_0\}) = 0$, hence $\nu(\{y\}) = 0$ for any $y
\in \gD$. This means that $\nu_s$ has to be singular continuous. Let
us introduce the function $\gth(t) := \nu_s([0,t))$, $t \in [0,t)$.
The function $\nu_s(t)$ is continuous and monotone. From $\nu'_s(y) =
0$ we get that the derivative of $\gth'(y)$ exists  and $\gth'(y) = 0$
for each $y \in \gD$. Hence the function is constant which yields that
$\nu_s(\gD) = 0$.

Conversely, let us assume that $\wt{h}(\cdot)$ is continuous on $\gD$,
$\nu_s(\gD) = 0$, and condition \eqref{5.47} holds. Then we have the
representation
\bead
\lefteqn{
|f(x+iy)| = |D(x+iy)|}\\
& &
\times \exp\left\{
-\frac{1}{\pi}\int_\bR\frac{x}{x^2 + (y-t)^2}\,d\nu_s(t)
- \frac{1}{\pi}\int_\bR\frac{x}{x^2 + (y-t)^2}\,\wt{h}(t)\,dt
\right\}e^{-ax}
\nonumber
\eead
If $y \in \gD$, then $\lim_{x\to+0}\frac{1}{\pi}\int_\bR\frac{x}{x^2 +
  (y-t)^2}\,d\nu_s(t) = 0$. Since $\wt{h}(\cdot)$ is continuous on the interval $\gD$ we
have $\lim_{x\to+0}\frac{1}{\pi}\int_\bR\frac{x}{x^2 +
  (y-t)^2}\,\wt{h}(t)\,dy = \wt{h}(y)$ for each $y \in \gD$. Thus we find
$\lim_{x\to+0}|f(x+iy)| = e^{-\wt{h}(y)}$ for each $y \in \gD$ and the
limit $|f|(iy)$ is continuous on $\gD$. Since $\wt{h}(y)$ is finite for each
$y \in \gD$ the limit $|f|(iy)$ is different from zero for each $y \in \gD$.
\end{proof}
Conditions of the type appearing in the proposition were discussed in
\cite{Mul84}. In particular, it turns out that the condition
\eqref{5.47} is satisfied if and only if
\be\la{5.51}
\lim_{x\to+0}\frac{\gt(iy,x)}{x} = 0
\ee
holds for every $y \in \gD$ where
\be\la{5.52}
\gt(iy,t) := \sum_{|iy - \xi_{k}|\le t}\real(\xi_k), \quad y \in \bR_+,
\quad t > 0.
\ee
It is clear that the validity of the condition \eqref{5.51} is related 
to the distribution of zeros in $\bC_\mathrm{right}$.
Of course, if there is only a finite number of zeros $\xi_k$, 
then condition \eqref{5.51} is satisfied. 
\bt\la{V.3}
Let $f(\cdot)$ is a holomorphic Kato function and let $\gD$ be an open
interval of $\bR$. The limit
$f(iy) = \lim_{x\to+0}f(x + iy)$ exists for every $y \in \gD$,
is locally H\"older continuous and different from zero on $\gD$ if and
only if the zeros of $f(\cdot)$ do not accumulate to any point of
$i\gD := \{iy: y \in \gD\}$, $\nu_s(\gD) = 0$ and the extended weight function 
$\wt{h} :=h(|t|)$, $t \in \bR$, is locally H\"older continuous on $\gD$.

In particular, the limit $f(iy)$ exists for
every $y \in \bR$, is locally H\"older continuous and different from zero on $\bR$ if
and only if $f(\cdot)$ has only a finite number of zeros in every
bounded open set of $\bC_{\rm
right}$, $\nu_s \equiv 0$ and the extended weight function $\wt{h}(\cdot)$ is locally H\"older
continuous on $\bR$.
\et
\begin{proof}
We note that the existence of the limit $f(iy) =
\lim_{x\to+0}f(x+iy)$ for each $y \in \gD$
yields the existence of $|f|(iy) = \lim_{x\to+0}|f(x+iy)|$
and the relation $|f(iy)| = |f|(iy)$ for each
$y \in \gD$. Hence $|f|(\cdot)$ is continuous. Applying Proposition \ref{V.2} we
get that condition \eqref{5.47} is satisfied, $\nu_s(\gD) = 0$ and
$\wt{h}(\cdot)$ is continuous. In fact, one has $h(y) = -\log(|f|(iy))$,
$y\in\gD$. This yields that the function $\wt{h}(\cdot)$ is locally
H\"older continuous on $\gD$ as well. If $\wt{h}(\cdot)$ is locally H\"older
continuous on $\gD$, then the limit
\bead
\lefteqn{
\varphi(y) := \lim_{x\to+0}\frac{i}{\pi}
\int_\bR\left(\frac{1}{iz-t} + \frac{t}{1+t^2}\right)d\nu(t) }\\
& &
\!\!\!\!\!\!\!\!=\lim_{x\to+0}
\left\{\frac{i}{\pi}\int_\bR\left(\frac{1}{iz-t}+\frac{t}{1+t^2}\right)d\nu_s(t)+
\frac{i}{\pi}\int_\bR\left(\frac{1}{iz-t} + \frac{t}{1+t^2}\right)\,\wt{h}(t)dt
\right\},
\nonumber
\eead
$z = x + iy \in \bC_\mathrm{right}$, exist for every $y \in \gD$. Indeed, we
have
\bea\la{5.55}
\lefteqn{
\frac{i}{\pi}\int_\bR\left(\frac{1}{iz-t} +\frac{t}{1+t^2}\right)d\nu_s(t) }\\
& &
=\frac{1}{\pi}\int_\bR \frac{x}{x^2 + (y-t)^2}\,d\nu_s(t) -
\frac{i}{\pi}\int_\bR \left(\frac{y-t}{x^2 + (y-t)^2} +
\frac{t}{1 + t^2}\right)d\nu_s(t)
\nonumber
\eea
where we have used $\nu_s(-X) = \nu_s(X)$. Taking into account that
$\nu_s(\gD) = 0$ we immediately get from
the representation \eqref{5.55} that the limit
\bed
\varphi_s(y) :=
\lim_{x\to+0}\frac{i}{\pi}\int_\bR\left(\frac{1}{iz-t}
  +\frac{t}{1+t^2}\right)d\nu_s(t) =
-\frac{i}{\pi}\int_\bR \frac{1+yt}{y-t}
\frac{d\nu_s(t)}{1 + t^2}\,,
\eed
$z = x + iy \in \bC_\mathrm{right}$, exist for each $y \in \gD$. Since
\bead
\lefteqn{
\frac{i}{\pi}\int_\bR\left(\frac{1}{iz-t} +\frac{t}{1+t^2}\right)\wt{h}(t)\,dt }\\
& &
=\frac{1}{\pi}\int_\bR \frac{x}{x^2 + (y-t)^2}\,\wt{h}(t)\,dt -
\frac{i}{\pi}\int_\bR \left(\frac{y-t}{x^2 + (y-t)^2} +
\frac{t}{1 + t^2}\right)\wt{h}(t)\,dt
\nonumber
\eead
we infer that
\bead
\lefteqn{
\varphi_{ac}(y) :=
\lim_{x\to+0}\frac{i}{\pi}\int_\bR\left(\frac{1}{iz-t} +\frac{t}{1+t^2}\right)\wt{h}(t)\,dt }\\
& &
=\wt{h}(y) +
\lim_{x\to+0}\frac{i}{\pi}\int_\bR \left(\frac{y-t}{x^2 + (y-t)^2} +
\frac{t}{1 + t^2}\right)\wt{h}(t)\,dt,
\nonumber
\eead
$z = x + iy \in \bC_\mathrm{right}$. If $\wt{h}(\cdot)$ is locally H\"older
continuous on $\gD$, then the limit
\bed
\wt{\varphi}_{ac}(y) := \lim_{x\to+0}\frac{i}{\pi}\int_\bR \left(\frac{y-t}{x^2 + (y-t)^2} +
\frac{t}{1 + t^2}\right)\wt{h}(t)dt
\eed
exists for each $y \in \gD$, and consequently, the limit 
$\varphi(y) = \varphi_s(y) + \varphi_{ac}(y)$ exist for every $y \in \gD$. Using the
representation
\be\la{5.63aa}
\exp\left\{\frac{i}{\pi}
\int_\bR\left(\frac{1}{iz-t} +
  \frac{t}{1+t^2}\right)\wt{h}(t)dt\right\}f(x+iy)e^{\ga z} =
D(x+iy)
\ee
for $z = x+iy \in \bC_\mathrm{right}$ we find the existence of the limit
\be\la{5.63}
D(iy) := \lim_{x\to+0}D(x+ iy)
\ee
for every $y \in \gD$. Taking into account \eqref{5.63aa} we find that
$D(iy)$ is continuous on $\gD$. Using the conformal mapping 
$\bC_{\rm right} \ni z \longrightarrow \frac{1-z}{1+z} \in
\bD := \{z\in \bC: |z| < 1\}$ which maps $\bC_{\rm right}$ onto $\bD$ and setting
\bed
B(z) := D((1-z)(1+z)^{-1}), \quad z \in \bD,
\eed
one defines a Blaschke product in $\bD$. The open set $\gD$ transforms
into an open set $\gd$ of $\bT := \{z\in\bC: |z| = 1\}$.  By the
Lindel\"of sectorial theorem \cite{Linde15} we get that $B(z)$
admits radial boundary values for each point of $\gd$.
The boundary function $B(e^{i\theta}) := \lim_{r\to 1}B(re^{i\theta})$
admits the representation
\be\la{5.63a}
B(e^{i\gth}) = D(-i\tan(\gth/2)), \quad e^{i\gth} \in \gd.
\ee
Since $D(iy)$ is continuous on $\gD$ the Blaschke product
$B(e^{i\gth})$ is continuous on $\gd$. If $e^{i\gth_0} \in \gd$ is an
accumulation point of zeros of, then for every $\epsilon > 0$ the set 
$\{B(e^{i\gth}: |\gth-\gth_0| < \epsilon\}$ contains $\bT$, see
\cite[Chapter 5]{ColLoh66} or \cite[Remark 4.A.3]{Col85}. Since
$B(e^{i\gth})$ is continuous on $\gd$, this is impossible which shows
that $e^{i\gth_0}$ is not an accumulation point of zeros of $B(z)$. Hence no point of
$\gd$ is an accumulation point which yields that no point of $\gD$ is
an accumulation point of zeros of $f(\cdot)$. 

Conversely, let us assume that no point of $i\gD$ is an
accumulation point of zeros of $f(\cdot)$. This yields that no point of
$\gd$ is an accumulation point of zeros of $B(z)$. Since 
$\inf_{k \in \bN}|e^{i\gth} - z_k| >0$ for any $e^{i\gth} \in \gd$
by a result of Frostman \cite{Fro42} one gets that the radial boundary
values $B(e^{i\gth}) = \lim_{r\to 1}B(re^{i\gth})$ exist for each
$e^{i\gth} \in \gd$. Using \cite[Remark 4.A.2]{Col85}
we get that $B(e^{i\gth})$ is continuous on $\gd$.  Applying again
the Lindel\"of sectorial theorem \cite{Linde15} we find that
$D(iy)$ exists for each $y \in \gD$ and is continuous. 

Since $\nu_s(\gD) = 0$ the limit
$\varphi_s(\cdot)$ exists for every $y \in \gD$. Because
$\wt{h}(\cdot)$ is locally H\"older continuous on $\gD$ we 
conclude that the limit $\varphi_{ac}(y)$ exist for every $y \in \gD$. 
Hence the limit $\varphi(y)$ exists for every $y \in \gD$ and 
\bed
S(iy) := \lim_{x\to+0}\exp\left\{-\frac{i}{\pi}
\int_\bR\left(\frac{1}{iz-t} +
  \frac{t}{1+t^2}\right)d\nu(t)\right\}e^{-\ga z},
\eed
$z = x+ iy \in \bC_\mathrm{right}$, exists for every $y \in \gD$. 
In this way we have demonstrated the existence of $f(iy)$ and the representation
$f(iy) = D(iy)S(iy)e^{-iay}$ for each $y \in \gD$. Using this representation
we get that $f(iy)$ is locally H\"older continuous on $\gD$ and different from zero.

If the limit $f(iy)$ exist for each $y \in \bR$, is locally H\"older continuous and different
from zero, then in view of the first part no point of the
imaginary axis is an accumulation point of zeros of $f(\cdot)$. Therefore,
any rectangle of the form $\kO := \{z \in
\bC_{\rm right}: |\imag(z)| < y_0, \quad 0 < \real(z) < x_0\}$
contains only a finite number of zeros. Otherwise, it would be exists an
imaginary accumulation point. Hence any bounded open sets contains only
a finite number of zeros. From the first part it follows that
$\wt{h}(\cdot)$ is locally H\"older continuous on $\bR$.

Conversely, if any open set contains only a finite
number of zeros, then, in particular, the rectangle of the form
$\kO$ contains only a finite number of zeros. Hence 
imaginary accumulation points do not exists.  By the first part 
it immediately follows that $f(\cdot)$ is locally H\"older continuous
and different from from zero on $\bR$. 
\end{proof}

\subsection{Examples}

\begin{enumerate}

\item If the holomorphic Kato function $f(\cdot)$ has no zeros in
  $\bC_\mathrm{right}$ and $\nu \equiv 0$, then $f(z) = e^{-z}$, $z \in
  \bC_\mathrm{right}$, where $\ga=1$ follows from condition \eqref{5.36}.

\item
If the holomorphic Kato function $f(\cdot)$ has zeros and
the measure $\nu \equiv 0$, then
$f(\cdot)$ is of the form $f(z) = D(z)e^{-\ga z}$,  where the Blaschke-type
product $D(z)$ is given by \eqref{5.26}. In particular, if $n =1$ we find the representation
\bed
f(z) = \frac{z^2 - 2z\real(\xi) + |\xi|^2}{z^2 + 2z\real(\xi) +
  |\xi|^2}\, e^{-\ga z},
\quad z \in \bC_\mathrm{right}\,,
\eed
where $\xi \in \bC_\mathrm{right}$ such that
\bed
\ga + 4\frac{\real(\xi)}{|\xi|^2} = 1.
\eed
This gives the representation
\be
f(z) =
\frac{z^2 -2\eta\left(z - \frac{2}{1-\ga}\right)}{z^2 + 2\eta
\left(z + \frac{2}{1-\ga}\right)}\, e^{-\ga z},
\quad z \in \bC_\mathrm{right},
\ee
$0 < \eta \le \frac{4}{1-\ga}$, $ 0 \le \ga \le 1$, where we have 
denoted $\xi = \eta + i\gt$, $\eta > 0$, and
$\gt = \sqrt{\frac{4}{(1-\ga)^2} - \left(\eta - \frac{2}{1-\ga}\right)^2}$.
The limit  $f(iy) := \lim_{\epsilon\to+0}f(\epsilon + iy)$, $y \in \bR$,
exists for each $y \ge 0$ and is given by
\bed
f(iy) = \frac{y^2 + 4\eta\frac{1}{1-\ga} + 2i\eta y}{y^2 -
  4\eta\frac{1}{1-\ga} +2i\eta y}e^{-i\ga y} =: \phi(y), \quad y \in \bR.
\eed
We note that $\phi(\cdot)$ is admissible.

\item
If the holomorphic Kato function $f(z)$ has no zeros and 
the measure $\nu$ is atomar, then $f(z)$ admits the representation
\bed
f(z) = \exp\left\{-\frac{2z}{\pi}\sum_l\frac{1}{z^2 +
    s^2_l}\nu(\{s_l\})\right\}\, e^{-\ga z},
\quad z \in \bC_\mathrm{right},
\eed
where $\{s_l\}_l$ the point where $\nu(\{s_l\}) \not= 0$.  In
the particular case when $d\nu(t) = c\gd(t-s)dt$, $s > 0$, we have
\bed
f(z) = \exp\left\{-\frac{2zc}{\pi}\frac{1}{z^2 + s^2}\right\}e^{-\ga z},
\eed
and $\ga + \frac{2c}{\pi}\frac{1}{s^2} = 1$ which yields $c =
\frac{1}{2}(1-\ga)\pi s^2$ and
\bed
f(z) := \exp\left\{-z(1-\ga)\frac{s^2}{z^2 + s^2}\right\}e^{-\ga z}
\eed
The limit
$f(iy) := \lim_{\epsilon\to+0}f(\epsilon + iy)$, $y \in \bR$, exists
for all $y \in \bR \setminus \{-s,s\}$ and is given by
\bed
f(iy) = \exp\left\{iy(1-\ga)\frac{s^2}{y^2-s^2}\right\}e^{-i\ga y} := \phi(y), \quad y
\in \bR \setminus \{-s,s\}.
\eed
The function $\phi(y)$ is admissible.

\item
If the holomorphic Kato function $f(z)$ has no zeros and the
measure  $\nu$ is absolutely continuous,
that is, $d\nu(t) = h(t)dt$, $h(t)(1+t^2)^{-1} \in L^1(\bR_+)$, then
$f(z)$ admits the representation
\bed
f(z) = \exp\left\{-\frac{2z}{\pi}\int^\infty_0 \frac{h(t)}{z^2 + t^2}
  dt\right\}\,e^{-\ga z},
\quad z \in \bC_\mathrm{right}
\eed
such that
\bed
\ga + \lim_{x\to+0}\frac{2}{\pi}\int^\infty_0 \frac{h(t)}{x^2 + t^2}dt
=1.
\eed
In particular, if $f(x) = (1 + \tfrac{x}{k})^{-k}$, $x \in \bR_+$,
then the holomorphic continuation $f(z) = (1 + \tfrac{z}{k})^{-k}$ has
no zeros which means that in the representation \eqref{5.5} the
Blaschke-type product $D(x)$ is absent. Moreover, the limit $f(iy) =
(1 + \tfrac{iy}{k})^{-k}$ exists for all $y \in \bR_+$, $|f(iy)|$ is
locally H\"older continuous and different from zero on $\bR_+$. 
Taking into account Theorem \ref{V.3} this yields the representation
\bed
f(z) =
\exp\left\{-\frac{kz}{\pi}
\int_{\bR_+}\frac{1}{z^2 + t^2}\ln\Big(1 +
\frac{t^2}{k^2}\Big)\;dt\right\}e^{-\ga z},
\quad z \in \bC_\mathrm{right}.
\eed
A straightforward computation shows that
\bed
\lim_{x\to+0}\frac{k}{\pi}\int_{\bR_+}\frac{1}{x^2 + t^2}\ln\Big(1 +
\frac{t^2}{k^2}\Big)\,dt = 1
\eed
which yields $\ga = 0$, and consequently, we have
\bed
f(z) = \exp\left\{-\frac{kz}{\pi}
\int_{\bR_+}\frac{1}{z^2 + t^2}\ln\Big(1 + \frac{t^2}{k^2}\Big)\,dt\right\}
\eed
for $z \in \bC_\mathrm{right}$.
\end{enumerate}

\subsection*{Acknowledgment}

The authors are grateful for the hospitality they enjoyed, 
P.E. in WIAS and H.N. in Doppler Institute, during the time when 
the work was done. The research was supported by the Czech Ministry of 
Education, Youth and Sports within the project LC06002.


\def\cprime{$'$} \def\cprime{$'$} \def\cprime{$'$} \def\cprime{$'$}
  \def\cprime{$'$} \def\cprime{$'$} \def\cprime{$'$} \def\cprime{$'$}
  \def\cprime{$'$} \def\lfhook#1{\setbox0=\hbox{#1}{\ooalign{\hidewidth
  \lower1.5ex\hbox{'}\hidewidth\crcr\unhbox0}}}

\end{document}